\documentclass[10pt,a4paper]{article}
\usepackage[T1]{fontenc}
\usepackage[utf8]{inputenc}
\usepackage{textcomp}
\usepackage[a4paper,textwidth=177mm,top=22mm,bottom=24mm]{geometry}
\usepackage{amsmath,amssymb,graphicx,booktabs}
\usepackage[authoryear,round]{natbib}
\usepackage{caption}
\usepackage[hidelinks,unicode]{hyperref}
\usepackage{url}
\newcommand{\introduction}{\section{Introduction}}
\newcommand{\conclusions}{\section{Conclusions}}
\newcommand{\codedataavailability}[1]{\section*{Code and data availability}#1}
\newcommand{\authorcontribution}[1]{\section*{Author contributions}#1}
\newcommand{\competinginterests}[1]{\section*{Competing interests}#1}
\newcommand{\videosupplement}[1]{\section*{Video supplement}#1}
\newenvironment{acknowledgements}{\section*{Acknowledgements}}{}
\date{}
\begin{document}
\title{Ionospheric bow waves during Starship Flight 14: coupling through vented propellant}
\author{Juha Vierinen\\[0.4em]
\small Department of Physics and Technology, UiT The Arctic University of Norway, Troms\o, Norway\\
\small \href{mailto:jvi019@uit.no}{jvi019@uit.no}}
\maketitle
\begin{abstract}
Global Navigation Satellite System (GNSS) observations of Starship Flight 14 show clear bow-shaped waves in detrended total electron content (dTEC), including during motor-off coast. These observations raise a coupling problem because the spacecraft hull is small compared with the thermospheric molecular mean free path. Sustained visible release in flight video motivates the hypothesis that vented, unburned propellant supplies a much larger interaction region capable of producing a fluid response. We model gaseous methane and oxygen using the Stochastic Parallel Rarefied-gas Time-accurate Analyzer (SPARTA) for direct simulation Monte Carlo calculations at a combined release rate of 10~kg~s$^{-1}$. The calculated collisional energy transfer extends over a region roughly 10~km across. We use this deposition pattern as a moving source of mass, momentum and total energy in a nonlinear neutral model coupled to a passive plasma response. The model produces a South American dTEC bow with broadly similar morphology and propagation speeds to the observations. Representative South American cuts have simulated root-mean-square (RMS) residuals of about 0.015 total electron content units (TECU), compared with observed values of 0.021--0.032 TECU; the North American response remains substantially weaker than observed. These results support collisional transfer of orbital momentum and energy by vented propellant as an explanation for the coast-phase waves. The release rate remains uncertain, and chemical contributions require further testing.
\end{abstract}

\introduction
Global Navigation Satellite System (GNSS) observations of Starship Flight 14 show clear bow-shaped waves in detrended total electron content (dTEC), including during coast when the rocket motors were off. The coast-phase waves pose a physical question: what couples the spacecraft to the atmosphere strongly enough to generate this large-scale disturbance? At thermospheric altitudes, the spacecraft hull is small compared with the ambient molecular mean free path and encounters free-molecular flow. Most spacecraft do not continuously release substantial quantities of gas, and comparable detectable, large-scale ionospheric bow waves are not a routine signature of their passage. If passage of the spacecraft body alone were sufficient to produce such waves, comparable signatures would be widespread. Starship's sustained venting therefore provides a physically distinct mechanism for coupling its orbital motion to the atmosphere.

Flight video provides a clue: sustained visible release of material during coast \citep{flight14webcast}. We interpret this as a controlled release of unburned propellant and model it as gaseous methane and oxygen. Gas released from a spacecraft initially carries approximately its orbital velocity relative to the atmosphere. As it expands and collides with ambient molecules, it can transfer momentum and energy over a region much larger than the hull. Our hypothesis is that this distributed transfer drives the neutral disturbance that appears as bow-shaped structure in dTEC. Combustion is not required for this mechanism.

Rocket activity can generate extensive ionospheric waves \citep{lin2014,lin2017,chou2018circular}. Moving exhaust sources in coupled atmosphere--ionosphere models produce V-shaped disturbances \citep{lin2017rocket}, and similar geometry was reported for the 18 November 2023 Starship flight \citep{yasyukevich2024starship}. The present event extends the coupling question to coast at a higher altitude, without sustained combustion exhaust.

We test whether vented propellant can supply that source. Direct simulation Monte Carlo (DSMC) calculations describe collisional transfer from released CH$_4$ and O$_2$ to the neutral atmosphere. Their spatial deposition pattern defines a moving three-dimensional source of mass, momentum and total energy in a compressible neutral model coupled to a passive plasma response. We compare the resulting dTEC maps and time--distance diagrams with the observations. The objective is to establish whether this mechanism can produce the observed bow morphology and propagation scale; the uncertain release rate is treated as a prescribed scenario.

\section{Data}
\begin{figure}[tbp]
\centering
\begin{minipage}[t]{0.49\textwidth}
\raggedright\textbf{(a)}\par\smallskip
\includegraphics[width=\linewidth]{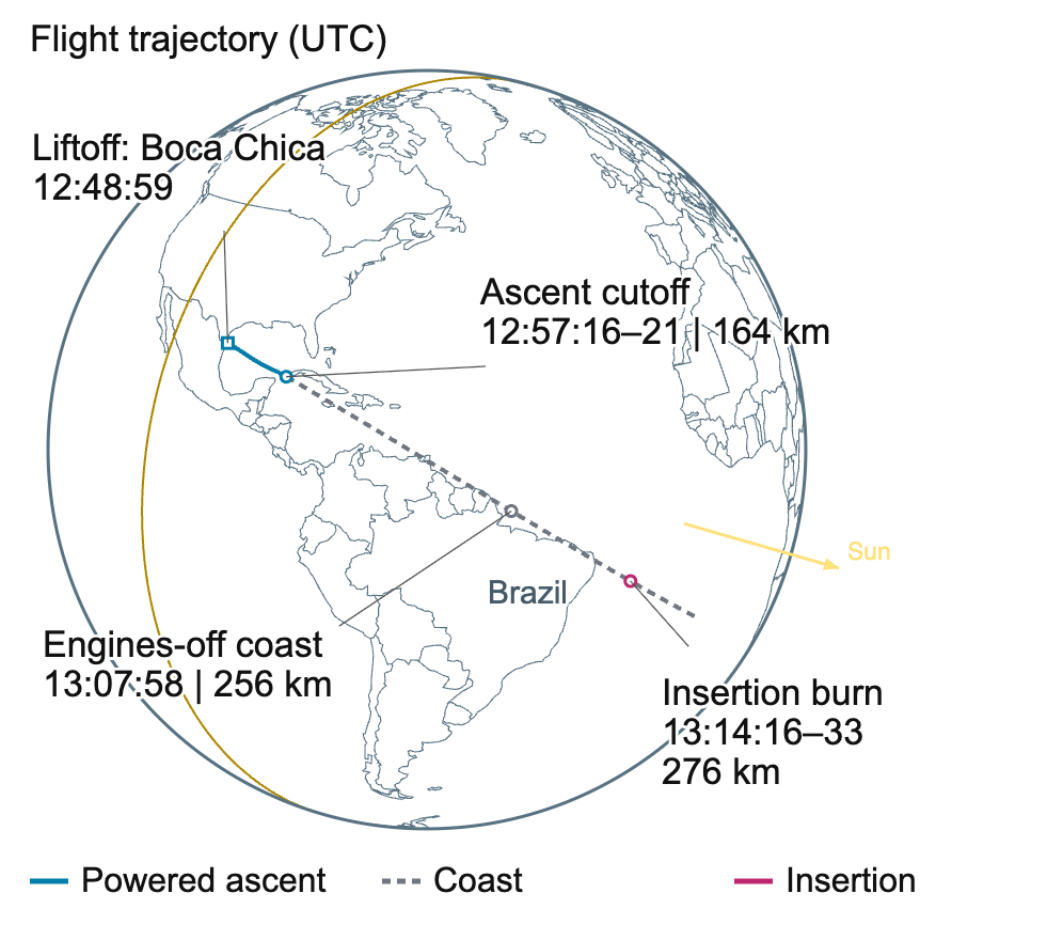}
\end{minipage}\hfill
\begin{minipage}[t]{0.49\textwidth}
\raggedright\textbf{(b)}\par\smallskip
\includegraphics[width=\linewidth]{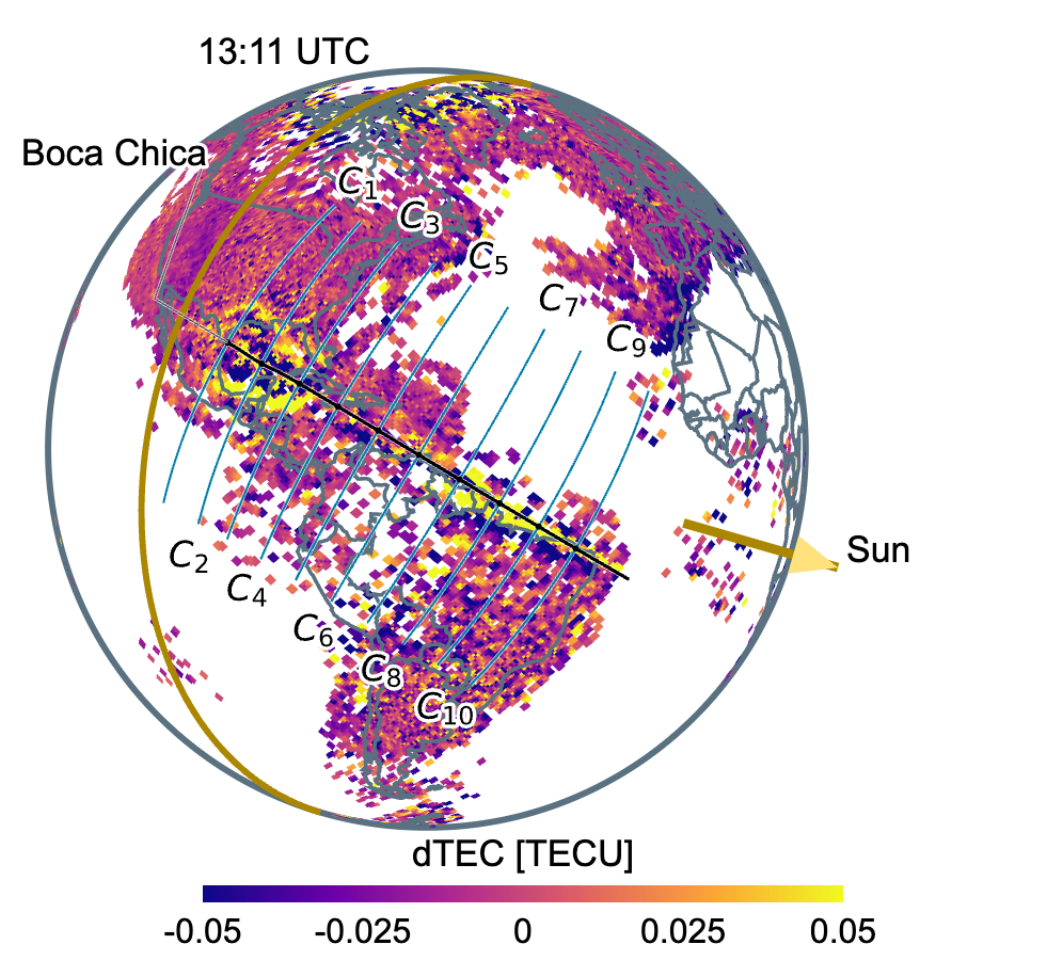}
\end{minipage}
\caption{Flight trajectory and keogram geometry on 28 September 2026. (a) Main flight phases during the first 30 min, with UTC and approximate altitude. Dashed grey marks motor-off coast over Brazil. The insertion engine turns on within 13:14:16--21 and off within 13:14:31--33, according to broadcast indicators. UTC uses the archived 12:48:59 liftoff anchor; altitudes are interpolated from the track. (b) Perpendicular cuts over the 13:11 UTC dTEC map. The black curve is the archived ground track. Blue cuts $C_1$--$C_{10}$ are spaced every 800 km from Boca Chica to 7200 km along track, with $C_1$ passing through the launch site and oriented perpendicular to the outbound bearing at approximately 100 km range. Labels alternate between north and south endpoints. Each cut extends 3000 km on either side; full corridor widths are 400 km for $C_1$--$C_2$ and 800 km for $C_3$--$C_{10}$. Its camera and colour scale match Fig.~\ref{fig:globe}. Gold marks the 13:11 UTC terminator and Sun direction in both panels.}
\label{fig:burn-map}
\label{fig:keogram-geometry}

\end{figure}

Starship Flight 14 launched from Starbase, Texas, on 28 September 2026 \citep{spacexflight}. We reconstruct the spacecraft trajectory from 489 archived position records from SpaceX's public tracker, preserved by StarDash \citep{stardash}. We use an archived liftoff time of 12:48:59 Coordinated Universal Time (UTC). Simulated and smoothed tracks are excluded. Broadcast engine indicators (Fig.~\ref{fig:burn-map}a) show ascent cutoff at about 12:57:16--12:57:21, coast during the first Brazilian passage at 13:07:58, and a brief insertion burn at about 13:14:16--13:14:33. The later deorbit burn occurs near 15:01:17--15:01:25, after the second South American pass, which produced a weaker but detectable dTEC perturbation \citep{flight14webcast}.

\begin{figure}[tbp]
\centering
\includegraphics[width=\textwidth]{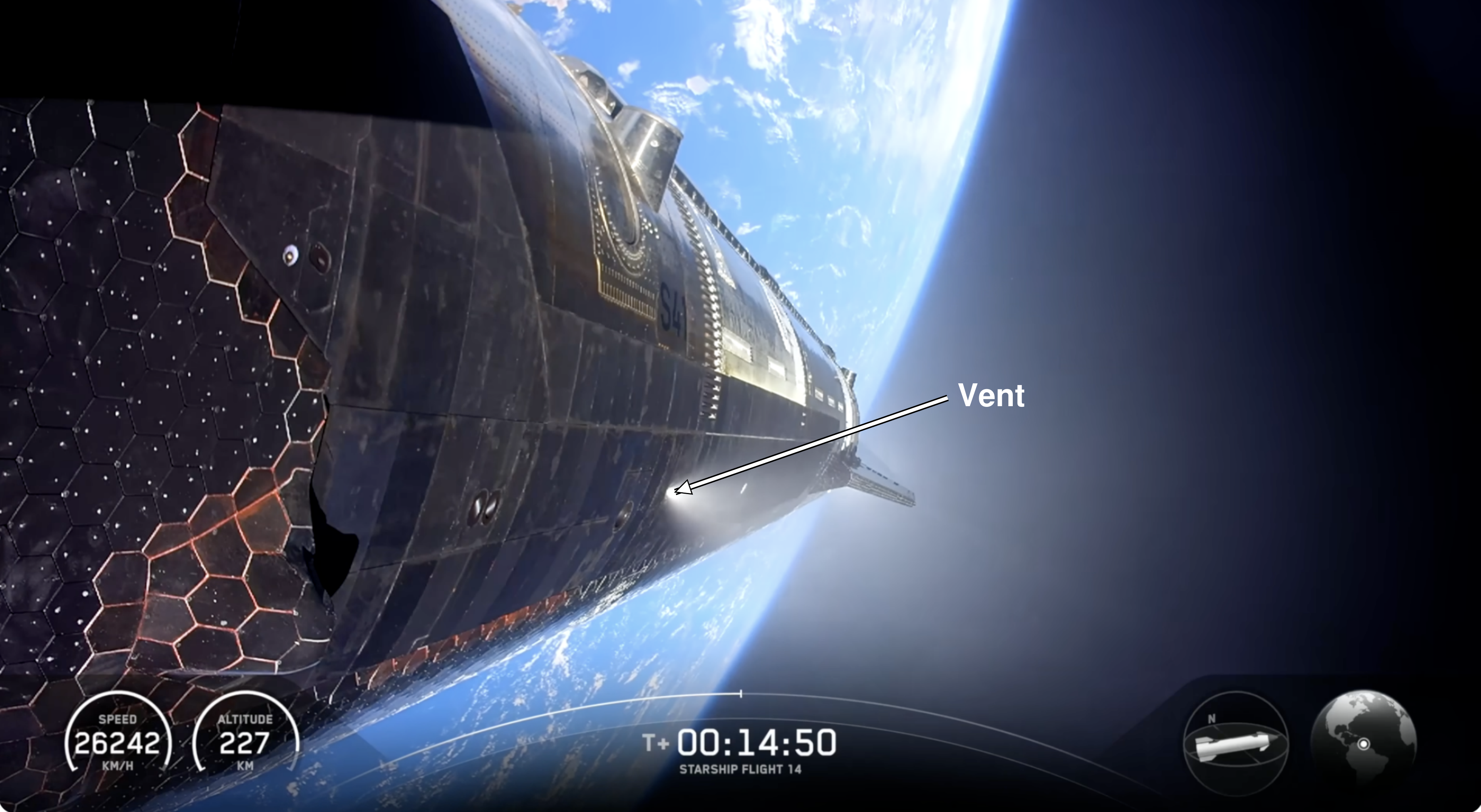}
\caption{Visible release from Starship during coast, at the displayed mission time T+00:14:50 and altitude 227 km. The arrow marks the apparent vent outlet at the origin of the bright plume. The image supports the presence of released material but does not determine its composition or mass flow. Frame from the Flight 14 video at 50:06 \citep{flight14ventvideo}.}
\label{fig:visible-vent}

\end{figure}

The video shows sustained visible release during coast (Fig.~\ref{fig:visible-vent}). We interpret it as unburned propellant and assume gaseous CH$_4$ and O$_2$ for modeling; the images do not measure composition or mass flow. Earlier Starship planning documents describe coast-phase propellant venting \citep{faastarship2024vent}. 

We obtain the ionospheric observations from the High-resolution Estimation of Ionospheric Maps through Differential Analysis of L-band Links (HEIMDALL) database \citep{heimdallAcknowledgement00}. We use 600~s detrended GNSS total electron content (TEC) at one-minute cadence, together with native receiver--satellite residuals and viewing geometry. The maps average dTEC fluctuations on a Hierarchical Equal Area isoLatitude Pixelation (HEALPix) grid \citep{gorski2005healpix} with resolution parameter $N_{\rm side}=128$, corresponding to an approximately 51~km surface cell scale.

\section{Methods}
\subsection{TEC maps and propagation}
Observed dTEC is obtained by subtracting a first-order Savitzky--Golay local-polynomial trend \citep{savitzky1964} from each continuous receiver--satellite TEC series, using a centred 600~s window. The linear least-squares fit uses the actual sample times. Windows require support on both sides and are interrupted at data gaps and quality boundaries. Observations above $15^{\circ}$ elevation are mapped in amplitude using a daily International Reference Ionosphere (IRI) profile and located on a 350~km pierce-point shell. Map cells are inverse-variance averages. One neighbouring ring of empty cells is filled for display only. All maps and keograms use $\pm0.05$ TECU colour limits, where one total electron content unit (TECU) is $10^{16}$ electrons m$^{-2}$.

Keograms show time against distance away from the ground track. We place ten perpendicular great-circle cuts every 800~km along track, from Boca Chica to 7200~km (Fig.~\ref{fig:keogram-geometry}b). Corridor widths are 400~km for $C_1$--$C_2$ and 800~km for $C_3$--$C_{10}$. Samples on both sides are pooled within each receiver in one-minute, 50~km bins, then receivers are averaged equally; bins require two receivers. Each panel has one reference slope: the sound speed in the modeled neutral atmosphere at the spacecraft altitude at cut passage, with a minimum reference altitude of 150~km. Sensitivity to mapping height and corridor width limits precise crest-speed inference.

\subsection{From molecular release to a distributed source}
We use DSMC to model the distribution of vented propellant and its transfer of momentum and energy to the surrounding atmosphere. The calculations use the Stochastic Parallel Rarefied-gas Time-accurate Analyzer (SPARTA) \citep{plimpton2019sparta}, with a combined methane and oxygen release of 10~kg~s$^{-1}$, an O$_2$:CH$_4$ mass ratio of 3.6:1, and a spacecraft speed of 7.6~km~s$^{-1}$ relative to the atmosphere.

The ambient density, temperature and composition represent conditions over Brazil at 265~km altitude and are supplied by the Naval Research Laboratory Mass Spectrometer and Incoherent Scatter model (NRLMSIS) version 2.1 \citep{emmert2021,emmert2022}. This model also supplies the reference atmosphere for the fluid calculation below. The three-dimensional DSMC calculation includes elastic collisions and excludes chemistry. Figure~\ref{fig:dsmc-wake} shows where energy is transferred to ambient gas, averaged over 4--12~s in the atmosphere frame.

\subsection{Atmospheric response and modeled dTEC}
We model the atmosphere as a three-dimensional compressible fluid. Vented propellant is added along the spacecraft trajectory as a moving source of mass, momentum and energy, producing changes in gas density, temperature and velocity. Collisions with the moving neutral gas drive an ionospheric plasma response, which changes the electron density and hence TEC. The coupling is one-way: the plasma responds to the neutral motion but does not affect it.

Away from the numerical boundary absorbers, the neutral model evolves mass, momentum and total energy through the compressible Navier--Stokes equations,
\begin{align}
 \partial_t\rho+\nabla\cdot(\rho\boldsymbol u)&=S_\rho,\\
 \partial_t(\rho\boldsymbol u)+\nabla\cdot(\rho\boldsymbol u\boldsymbol u+p\boldsymbol I)
 &=\rho\boldsymbol g-2\rho\boldsymbol\Omega\times\boldsymbol u
 +\nabla\cdot\boldsymbol\tau+\boldsymbol S_m+\boldsymbol F_{\rm eng},\\
 \partial_t E+\nabla\cdot[(E+p)\boldsymbol u]
 &=\rho\boldsymbol g\cdot\boldsymbol u
 +\nabla\cdot(\boldsymbol\tau\cdot\boldsymbol u-\boldsymbol q)
 +S_E+Q_{\rm eng}.
 \label{eq:neutral-energy}
\end{align}
Here $\rho$, $\boldsymbol u$ and $p$ are neutral mass density, velocity and pressure, $E=\rho e+\rho|\boldsymbol u|^2/2$ is total energy density, and $e$ is specific internal energy. The ideal-gas pressure uses the local mixture composition; injected methane and oxygen are transported with the flow. The identity tensor is $\boldsymbol I$, $\boldsymbol g$ is gravitational acceleration and $\boldsymbol\Omega$ is Earth's rotation vector. Viscous stress $\boldsymbol\tau$ and heat flux $\boldsymbol q$ redistribute momentum and energy; conduction acts on temperature perturbations to preserve the reference atmosphere. The engine terms are the applied force per volume $\boldsymbol F_{\rm eng}$ and total energy input $Q_{\rm eng}=Q_{\rm heat}+\boldsymbol F_{\rm eng}\cdot\boldsymbol u$, comprising prescribed heating and work done by that force.

The source follows a spatial weighting function $K$, which moves with the spacecraft and satisfies $\int K\,\mathrm dV=1$. Its shape is based on the positive atmosphere-frame energy transfer calculated by DSMC (Fig.~\ref{fig:dsmc-wake}), averaged around the flight axis. The prescribed release rate sets the total input, while $K$ distributes it through space.

For atmosphere-frame injection velocity $\boldsymbol v_s$ and combined mass rate $\dot m$, the volumetric sources are
\begin{equation}
 S_\rho=\dot mK,\qquad
 \boldsymbol S_m=\dot m\boldsymbol v_sK,\qquad
 S_E=\tfrac12\dot m|\boldsymbol v_s|^2K.
 \label{eq:vent-conservative-sources}
\end{equation}
We add mass, momentum and total energy together. After the source update, the bulk kinetic energy is calculated from the updated mass and momentum and subtracted from the updated total energy; the remainder is internal energy. Thus the partition between bulk motion and heating follows from the mixing update, rather than an independently prescribed heating fraction. The DSMC energy-transfer pattern defines $K$ and is not added again as heat. Using the same pattern for all three inputs assumes mixing of the released gas within this region. Initial thermal and ship-relative ejection energies are neglected.

The domain covers $35^{\circ}$ S--$55^{\circ}$ N, $115^{\circ}$ W--$15^{\circ}$ W and 0--1000~km altitude. A resting, altitude-dependent Boca Chica NRLMSIS profile is prescribed throughout. The finite-volume grid has $0.5^{\circ}$ horizontal spacing and 149 layers, with vertical spacing increasing from 2.5 to 25~km.

Continuous 10~kg~s$^{-1}$ venting starts 502~s after launch and follows the archived trajectory. Powered-flight forcing uses prescribed heat and momentum coupling fractions of 10\% during ascent and 1\% during insertion. These fractions and the vent rate are scenarios, not fitted flight parameters. The calculation ends at 14:50 UTC.

The ionospheric model represents singly ionized oxygen (O$^+$). Neutral collisions accelerate the ions, while the magnetic field deflects their motion and pressure gradients redistribute the plasma. The ion velocity and density evolve as
\begin{align}
 (\partial_t+\boldsymbol v_i\cdot\nabla)\boldsymbol v_i
 &=\nu_{in}(\boldsymbol u-\boldsymbol v_i)
 +\Omega_i\boldsymbol v_i\times\boldsymbol b
 -c_i^2\nabla\ln(N/N_0),\\
 \partial_tN+\nabla\cdot(N\boldsymbol v_i)&=0.
 \label{eq:plasma-continuity}
\end{align}
Here $\boldsymbol v_i$ is ion velocity, $N=N_0+\delta N_e$ is the common ion and electron number density, and $N_0$ is the fixed IRI background. The collision frequency is $\nu_{in}$, $\boldsymbol b$ is the magnetic-field unit vector, $\Omega_i$ is the ion gyrofrequency, and $c_i$ is the prescribed ion-acoustic speed. The field is supplied by the International Geomagnetic Reference Field (IGRF); temperatures and collision rates are prescribed and held fixed. The pressure term preserves $N=N_0$ as an equilibrium. Chemistry and evolving electric fields are omitted.

Boundary absorption is applied separately from these interior equations. Near the lateral and upper boundaries, neutral variables relax toward the resting reference atmosphere, while ion velocities and electron-density perturbations are damped toward zero. The plasma also has an absorbing layer near its lower boundary at 100~km. These numerical layers reduce artificial wave reflections; they do not represent physical atmospheric losses. The neutral ground boundary reflects vertical flow.

We integrate the electron-density perturbation through the complete modeled column to obtain the vertical total electron content perturbation ($\Delta\mathrm{VTEC}$),
\begin{equation}
 \Delta\mathrm{VTEC}(\lambda,\phi,t)=10^{-16}
 \int_{100\,\mathrm{km}}^{1000\,\mathrm{km}}\delta N_e(\lambda,\phi,z,t)\,\mathrm dz,
\end{equation}
then apply the same centred 600~s detrending as for the observations. Model keograms use the observed cut corridors and bins, but average both sides without a receiver mask or amplitude adjustment. Thus the model and observations retain different viewing geometries. The 14:47 map alone uses a shifted 14:40--14:50 fit because the simulation ends before a full centred window is available.

\section{Results}
\label{sec:results}
\subsection{Bow-shaped waves during motor-off coast}
\begin{figure}[tbp]
\centering\includegraphics[width=\textwidth]{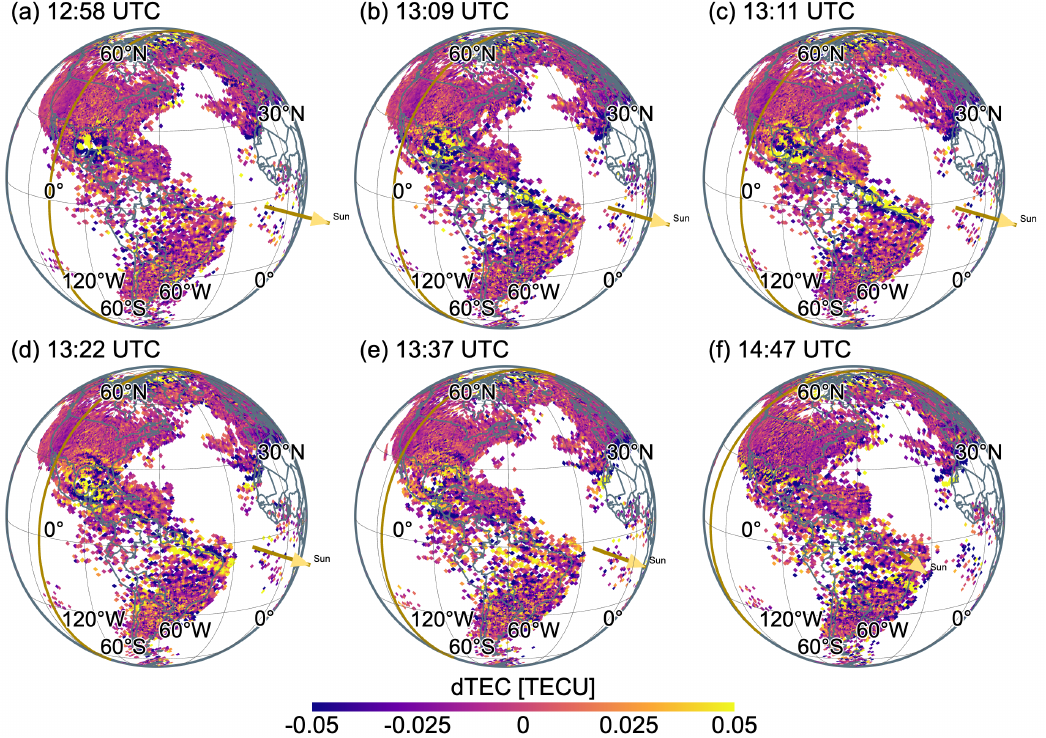}
\caption{Observed 600~s detrended TEC on 28 September 2026. The South American bow is visible during coast at 13:09 and 13:11 UTC and recurs during the later passage at 14:47. North American curved fronts are also apparent. All panels share a camera and $\pm0.05$ TECU colour scale. Labels denote one-minute bin beginnings. Grey lines mark geographic coordinates; gold marks the terminator and Sun direction. A single ring of neighbouring empty pixels is filled for display.}
\label{fig:globe}

\end{figure}

Figure~\ref{fig:globe} shows two distinct parts of the observed response: curved fronts expanding across North America and a narrow, bow-shaped disturbance aligned with the spacecraft's passage across South America. The South American bow is clear at 13:09 and 13:11 UTC, during the motor-off coast interval and before the short insertion burn. Its alternating positive and negative dTEC bands spread away from the trajectory as the spacecraft moves onward. Thus sustained engine firing over South America is not required for the observed bow waves.

A renewed track-aligned disturbance is visible at 14:47 UTC during the next South American passage. The nearest archived position, at 14:47:13, is $16.47^{\circ}$ S, $40.41^{\circ}$ W and 275.7~km altitude. This signature precedes the deorbit burn by about 14~min, longer than the forward support of the centred detrending window. Both passages therefore motivate a source associated with coast-phase flight.

\subsection{A deposition region much larger than the spacecraft}
\begin{figure}[tbp]
\centering

\begingroup
\normalsize
\dimen0=\linewidth
\advance\dimen0 by -177.000000mm
\ifdim\dimen0<0pt \dimen0=-\dimen0\fi
\ifdim\dimen0>0.1mm
  \PackageError{rustplotlib}{Article linewidth changed}{Regenerate this figure using the new ArticleLayout.}
\fi
\dimen0=\csname f@size\endcsname pt
\advance\dimen0 by -10.000000pt
\ifdim\dimen0<0pt \dimen0=-\dimen0\fi
\ifdim\dimen0>0.01pt
  \PackageError{rustplotlib}{Article body font changed}{Regenerate this figure using the new ArticleLayout.}
\fi
\includegraphics[width=168.470676mm]{\detokenize{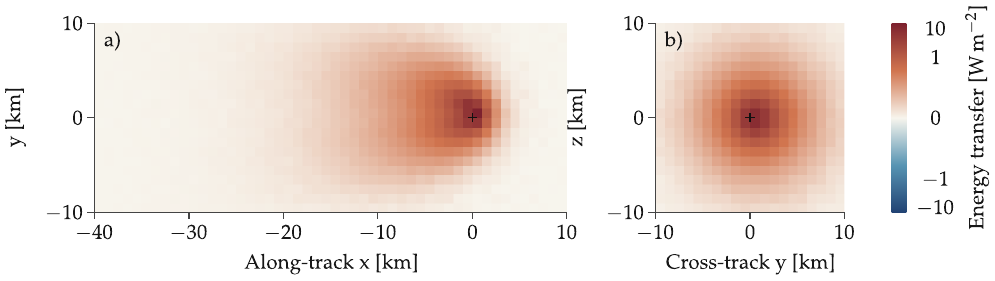}}
\endgroup

\caption{DSMC collision-mediated energy transfer for combined CH$_4$+O$_2$ release of 10~kg~s$^{-1}$. Panels integrate the atmosphere-frame ambient molecular energy transfer (a) over $z$ and (b) over $x$, averaged over 4--12~s. Coordinates are along track $x$, side-vent direction $y$ and vertical $z$; crosses mark release. The signed inverse-hyperbolic-sine colour scale shows W~m$^{-2}$. The transfer includes bulk and random molecular motion.}
\label{fig:dsmc-wake}

\end{figure}

The 10~kg~s$^{-1}$ DSMC calculation produces an extended collisional-transfer region behind and around the release point (Fig.~\ref{fig:dsmc-wake}). A transverse diameter of 11.4~km encloses 90\% of positive atmosphere-frame energy transfer over 8--12~s: a deposition region roughly 10~km across, compared with a 9~m hull. The ambient mean free path for the prescribed atmosphere and collision cross section is 1.94~km. Released gas thus spreads the transfer over a region much larger than the free-molecular spacecraft.

\subsection{Modeled bow waves and the observed propagation scale}
\begin{figure}[tbp]
\centering\includegraphics[width=\textwidth]{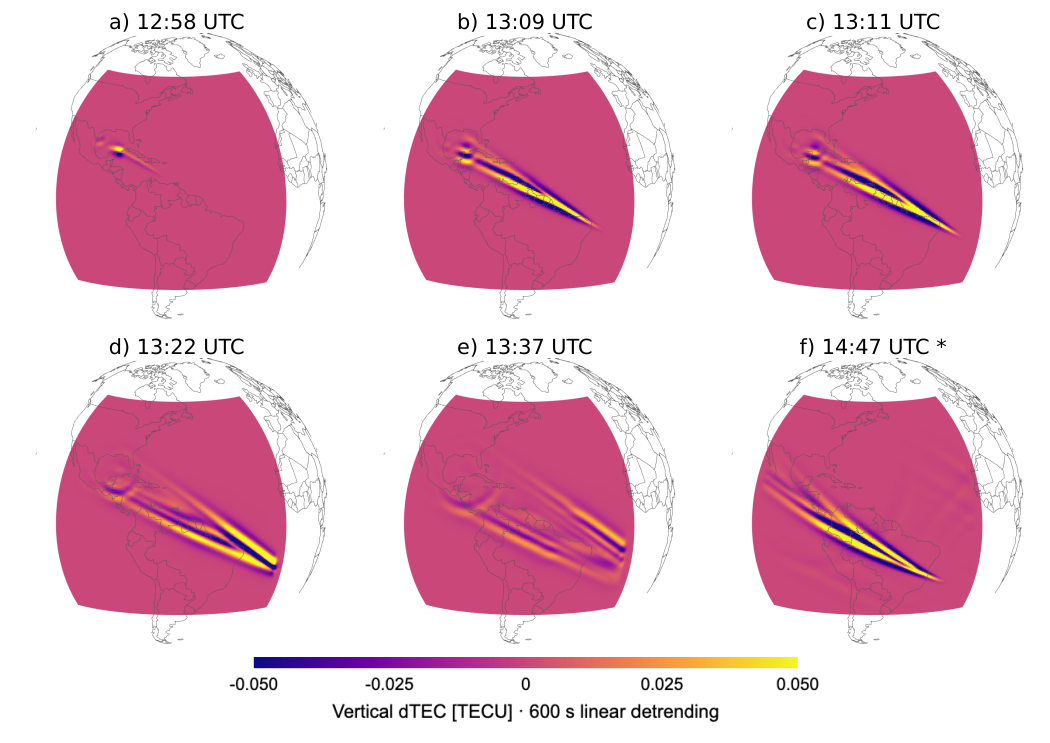}
\caption{Simulated vertical dTEC for the 10~kg~s$^{-1}$ combined CH$_4$+O$_2$ release scenario at the observation times in Fig.~\ref{fig:globe}. The calculation includes the prescribed ascent and insertion sources, neutral viscosity and thermal conduction. Each value integrates the 100--1000~km electron column before detrending. The full model domain is shown without GNSS sampling or coverage filling, on the same $\pm0.05$ TECU scale as the observations. Panels (a--e) use centred 600~s detrending; the asterisk in (f) identifies a shifted 14:40--14:50 fit.}
\label{fig:current-virtual-dtec}

\end{figure}

The 10~kg~s$^{-1}$ venting scenario generates an elongated bow-shaped dTEC response over South America (Fig.~\ref{fig:current-virtual-dtec}). Alternating-sign bands trail the moving source and broaden away from the ground track. This reproduces the principal organization of the observed coast-phase disturbance. The model also produces renewed track-aligned structure on the second passage, although the shifted filter at 14:47 makes that panel a qualitative comparison.

\begin{figure}[tbp]
\centering

\begingroup
\normalsize
\dimen0=\linewidth
\advance\dimen0 by -177.000000mm
\ifdim\dimen0<0pt \dimen0=-\dimen0\fi
\ifdim\dimen0>0.1mm
  \PackageError{rustplotlib}{Article linewidth changed}{Regenerate this figure using the new ArticleLayout.}
\fi
\dimen0=\csname f@size\endcsname pt
\advance\dimen0 by -10.000000pt
\ifdim\dimen0<0pt \dimen0=-\dimen0\fi
\ifdim\dimen0>0.01pt
  \PackageError{rustplotlib}{Article body font changed}{Regenerate this figure using the new ArticleLayout.}
\fi
\includegraphics[width=177.000000mm]{\detokenize{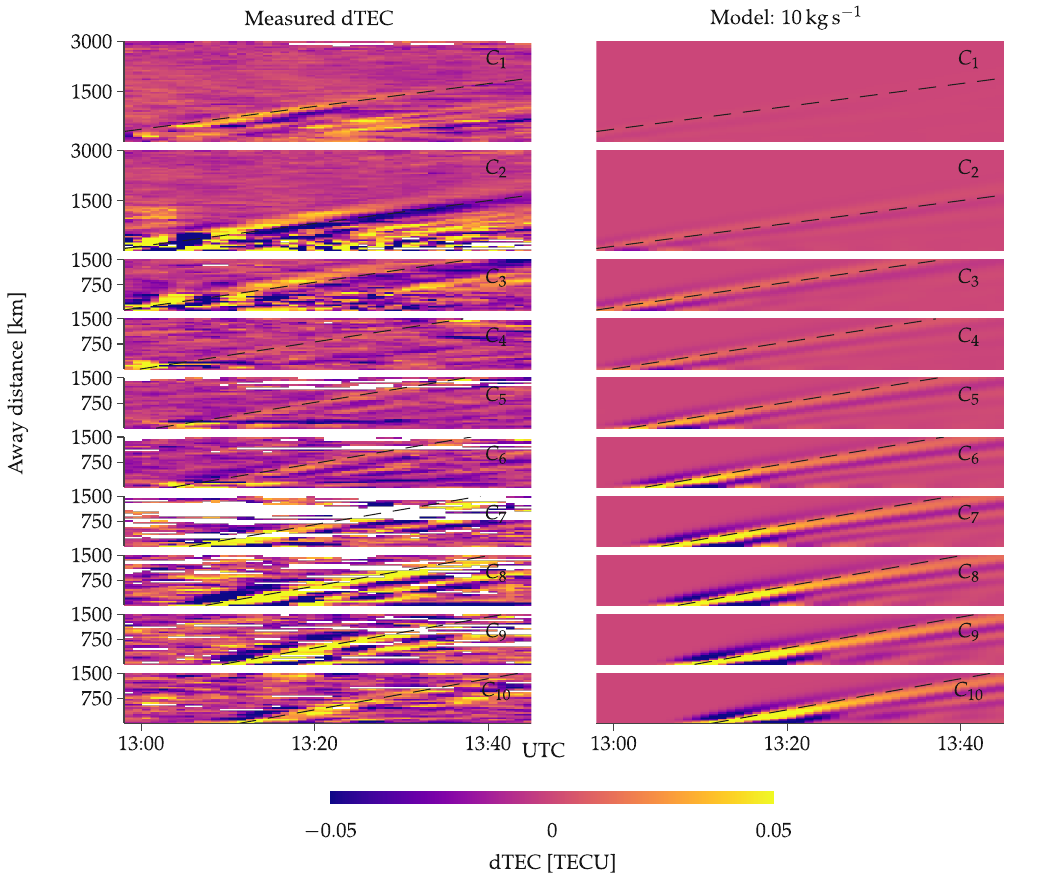}}
\endgroup

\caption{Measured dTEC (left) and the 10~kg~s$^{-1}$ modeled vertical dTEC (right) along all ten cuts in Fig.~\ref{fig:keogram-geometry}b. Panel heights preserve the distance--time scale. One dashed guide in each panel shows the modeled sound speed at the spacecraft altitude at cut passage, evaluated at no less than 150~km. Both fields use centred 600~s detrending and 50~km, one-minute bins; blank observations lack two receivers. Colours share $\pm0.05$ TECU limits.}
\label{fig:mechanical-paired-cuts}

\end{figure}

The paired keograms in Fig.~\ref{fig:mechanical-paired-cuts} use this same 10~kg~s$^{-1}$ calculation. In the South American cuts, both observations and model show outward-moving bands with comparable slopes on the scale of several hundred metres per second. The dashed guides show the modeled sound speed at spacecraft altitude, with a 150~km altitude floor: 567--774~m~s$^{-1}$ across the ten cuts. They provide a physical reference rather than fitted crest velocities. The spacecraft's locally derived Earth-fixed speed near 13:12 is 7.24~km~s$^{-1}$, well above these acoustic reference speeds.

The South American simulated residuals also reach a comparable amplitude scale. On common finite bins from 12:58 to 13:45 UTC and 150--1500~km from the track, root-mean-square (RMS) residuals are 0.032 and 0.021 TECU in observed $C_9$ and $C_{10}$, versus 0.015 and 0.015 TECU in the model. No amplitude multiplier is applied. The comparison is less successful over North America: in $C_1$ and $C_2$ over 150--3000~km, observed RMS residuals are 0.012 and 0.032 TECU, while the model gives 0.00037 and 0.0016 TECU. The modeled ascent response is therefore substantially weaker than observed. These whole-cut RMS values include all retained structure and do not isolate the amplitude of a matched wave crest.

The maps and keograms establish agreement in broad morphology and propagation scale. Individual crest associations remain sensitive to corridor width and assumed mapping height, so the comparison does not determine a unique wave altitude or propagation speed.

\section{Discussion}
The comparison supports venting as an explanation for the motor-off bow waves. Released gas extends the interaction far beyond the free-molecular hull, allowing collisions to transfer orbital momentum and energy to the thermosphere. The resulting moving source produces the observed broad South American morphology, comparable propagation slopes and dTEC amplitudes of the same order. Combustion is unnecessary: the energy reservoir is the motion of the released material relative to the atmosphere. A non-venting spacecraft lacks this extended material source.

The calculation demonstrates a sufficient candidate mechanism under prescribed conditions, rather than a unique reconstruction of the event. The 10~kg~s$^{-1}$ rate is uncertain, and the weak North American response shows that the model does not explain every observed feature quantitatively. The common deposition shape and eventual mixing are assumptions; full-event spatial convergence remains unestablished. Winds, source history and plasma conditions can change the response. Vertical model columns and GNSS sight lines sample different volumes, with possible phase cancellation \citep{yehliu1974}; consistent sight-line calculations would strengthen the amplitude test \citep{ducourtieux2022}. Acoustic and internal-gravity responses are both possible \citep{hines1960}, and the bow shape alone does not identify a unique mode or prove a nonlinear shock.

Mechanical sufficiency does not exclude chemistry. CH$_4$ and O$_2$ can participate in ion--molecule reactions that modify ion composition and electron loss \citep{bernhardt1987chemicals,cunhademiranda2015,bowden2020depletions}. Testing this contribution requires reactant concentrations and collision energies, not only an energy-deposition kernel. Mechanical-only, chemical-only and combined calculations should be compared for the same release history, retaining background TEC as well as detrended residuals. Independent modeling, including the Global Ionosphere--Thermosphere Model (GITM), can test morphology, propagation and amplitude with different transport and plasma formulations. That comparison remains future work.

\conclusions
Starship Flight 14 produced clear bow-shaped dTEC waves during motor-off coast, including a renewed disturbance during its next South American passage. Visible propellant release suggests a coupling mechanism that extends beyond the small spacecraft hull: expanding gas transfers orbital momentum and energy collisionally to the thermosphere.

A 10~kg~s$^{-1}$ DSMC calculation supports a deposition region roughly 10~km across. A conservative moving source representing this distributed coupling at the same prescribed combined CH$_4$+O$_2$ rate generates a bow-shaped dTEC response with broadly similar South American morphology and propagation speeds. Its residual amplitudes are of the same order as observed, although the North American response remains too weak.

Venting thus provides a physical explanation for otherwise unexpected coast-phase bow waves. The calculation establishes a plausible mechanical pathway without measuring the actual vent rate or excluding chemistry.

\codedataavailability{The neutral--plasma simulation and DSMC setup, documentation and input links are available at \url{https://github.com/jvierine/rns/tree/main/examples/starship14}. Instructions for accessing the simulation outputs will be added to that repository. Observations can be explored at \url{https://pikos.org/heimdall/gallery.html\#article-12}. SPARTA is available at \url{https://sparta.github.io/}.}
\authorcontribution{Juha Vierinen designed the study, performed the analysis and simulations, and prepared the manuscript.}
\competinginterests{The author declares that there are no competing interests.}
\videosupplement{The complete event movie at \url{https://youtu.be/Im9tYUsUOr8} presents measured HEIMDALL dTEC, modelled neutral mass-density perturbations at 300~km and modelled whole-column vertical dTEC, repeated three times. The observed movie uses HEIMDALL neighbouring-cell filling for display only. The observations span 11:49--14:55 UTC and include the second passage around 14:47 UTC; the model output ends at 14:50 UTC. Late model dTEC frames use the labelled shifted 14:40--14:50 UTC detrending window, while earlier frames use complete centred 600~s windows.}

\begin{acknowledgements}

We acknowledge HEIMDALL (High-resolution Estimation of Ionospheric Maps through Differential Analysis of L-band Links), the GNSS ionospheric data aggregator operated by UiT The Arctic University of Norway, Tromsø, Norway (\url{https://pikos.org/heimdall/}), the International Meridian Circle Program (IMCP) and its contributing institutions for supporting GNSS and ionospheric TEC data sharing, the Chinese Meridian Project, and the receiver operators, GNSS networks and archive services whose contributions are identified in the data-product provenance: the Finnish Meteorological Institute (FMI) / Geotrim (Finland); the Construction Sector Development Agency (SSVA) / Lithuanian Positioning System (LitPOS) (Lithuania); Instituto Geográfico Nacional (IGN) Argentina and its Red Argentina de Monitoreo Satelital Continuo (RAMSAC); Centro Nacional de Investigaciones Sismológicas (CENAIS) and GFZ for the SCUB station in Cuba, with archives provided by the Federal Agency for Cartography and Geodesy (BKG) and Système d’Observation du Niveau des Eaux Littorales (SONEL); the National Oceanic and Atmospheric Administration (NOAA) National Continuously Operating Reference Stations (CORS) Network; the BKG International GNSS Service (IGS) Data Center and Germany’s integrated geodetic reference network (GREF); Geobasis North Rhine-Westphalia (NRW) and its satellite positioning service (SAPOS); Instituto Geográfico Nacional (IGN) Spain and its GNSS reference-station network (ERGNSS); Institut national de l’information géographique et forestière (IGN) France and its Réseau GNSS permanent (RGP); the European Reference Frame (EUREF) Permanent GNSS Network; the French European Plate Observing System (EPOS-France) research infrastructure and Réseau national GNSS permanent (Rénag) network, with data distributed by the Rénag data centre; Istituto Nazionale di Geofisica e Vulcanologia (INGV) and its Rete Integrata Nazionale GPS (RING), where GPS denotes the Global Positioning System; Istituto Nazionale di Oceanografia e di Geofisica Sperimentale (OGS) and its Friuli Regional Deformation Network (FReDNet); Delft University of Technology (TU Delft) Dutch Permanent GNSS Array; SWEPOS, Sweden’s national permanent GNSS reference network; the Japan Aerospace Exploration Agency (JAXA) Multi-GNSS Monitoring Network (MGM-Net); the Hong Kong Satellite Positioning Reference Station Network (SatRef); the Netherlands Partnership Geodetic Infrastructure (NSGI) / Kadaster (Netherlands); Iceland’s continuously operating reference-station network (IceCORS) and the Icelandic Meteorological Office (Veðurstofa Íslands); Direção-Geral do Território (DGT) and its Rede Nacional de Estações Permanentes (ReNEP) (Portugal); Instituto Brasileiro de Geografia e Estatística (IBGE) and its Rede Brasileira de Monitoramento Contínuo dos Sistemas GNSS (RBMC) (Brazil); Instituto Geográfico Militar (IGM) Uruguay and its Red Geodésica Nacional Activa (REGNA-ROU); the University of California San Diego (UCSD) Scripps Orbit and Permanent Array Center (SOPAC) / Garner; the Royal Observatory of Belgium; Universidad de Santiago de Chile (USACH); Geoscience Australia’s Asia-Pacific Reference Frame (APREF); the New Zealand GeoNet programme and its sponsors the Natural Hazards Commission (NHC), Earth Sciences New Zealand, Land Information New Zealand (LINZ), the National Emergency Management Agency (NEMA) and the Ministry of Business, Innovation and Employment (MBIE); the National Science Foundation (NSF) National Geophysical Facility data archive operated by EarthScope Consortium (NSF award 2435260); the European Plate Observing System (EPOS) GNSS Data Gateway; Natural Resources Canada / Canadian Geodetic Survey and the Canadian Active Control System (CACS) network operators, with observations supplied under the Open Government Licence -- Canada; Québec’s Ministère des Ressources naturelles et des Forêts (MRNF) and its GNSS network operators; the University of New Brunswick Radio and Space Physics Laboratory, the Canadian High Arctic Ionospheric Network (CHAIN; Jayachandran et al., 2009, doi:10.1029/2008RS004046) and EclipseNB (Kashcheyev et al., 2025, doi:10.1029/2024SW004194), through \url{https://www.chain-project.net/data/gnss/data/}; Instituto Nacional de Estadística y Geografía (INEGI) and the Mexican Red Geodesica Nacional Activa (RGNA); Instituto Geofisico del Peru (IGP), Institut des Sciences de la Terre (ISTerre), California Institute of Technology (Caltech), Instituto Geografico Nacional del Peru (IGN) and Observatoire des Sciences de l’Univers de Grenoble (OSUG) for the historical Peruvian archives (doi:10.1029/2023JB027114); and the International GNSS Service, the Center for Orbit Determination in Europe (CODE) and the orbit-product providers identified in the provenance. HEIMDALL processes raw GNSS observations into derived TEC products. EarthScope and Rénag observations are provided under the Creative Commons Attribution 4.0 licence (CC BY 4.0) (\url{https://creativecommons.org/licenses/by/4.0/}); CACS observations are provided under the Open Government Licence -- Canada (\url{https://open.canada.ca/en/open-government-licence-canada}).
\nocite{heimdallAcknowledgement00,heimdallAcknowledgement01,heimdallAcknowledgement02,heimdallAcknowledgement03,heimdallAcknowledgement04,heimdallAcknowledgement05,heimdallAcknowledgement06,heimdallAcknowledgement07,heimdallAcknowledgement08,heimdallAcknowledgement09,heimdallAcknowledgement10,heimdallAcknowledgement11,heimdallAcknowledgement12,heimdallAcknowledgement13}

OpenAI Codex assisted editing and analysis-code preparation; the author retains responsibility for scientific interpretation.
\end{acknowledgements}
\bibliographystyle{copernicus}\bibliography{references}

\end{document}